\documentclass[prd,twocolumn,showpacs,preprintnumbers,aps,nofootinbib]{revtex4-1}

\usepackage{graphicx}
\usepackage{dcolumn}
\usepackage{bm}
\usepackage{amsmath,amssymb,amsfonts}
\usepackage{latexsym}

\usepackage{color}

\def\nn{\nonumber}

\def\l{\lambda}
\def\r{\rho}
\def\g{\gamma}

\def\rmm{\rho_{\mu\mu}}
\def\rtata{\rho_{\tau\tau}}

\def\rtt{\rho_{tt}}

\def\cg{c_\gamma}

\def\o{{\cal O}}
\begin{document}


\title{\boldmath
Charged Lepton EDM with Extra Yukawa Couplings
}

\author{Wei-Shu Hou, Girish Kumar and Sven Teunissen}
\affiliation{
Department of Physics, National Taiwan University, Taipei 10617, Taiwan}
\bigskip


\begin{abstract} 
In a two Higgs doublet model with extra Yukawa couplings,
we assess the new physics contributions to 
charged lepton electric dipole moment.
We focus especially on muon (and tau) EDM
where in the coming decade several experiments --- Muon g-2,
J-PARC and PSI (and Belle\;II) --- will push sensitivities
down by several orders of magnitude. 
With the working assumption that extra Yukawa couplings 
are analogous to SM ones in strength and taking 
exotic scalar masses in sub-TeV range, we find that
$\mu$ and $\tau$ EDM can be enhanced to values larger 
than new physics scenarios that scale with lepton mass.
The main effect comes from the flavor-conserving
extra top coupling $\rho_{tt}$ via well-known two-loop diagrams.
Deviating from our working assumption, 
if the muon $g-2$ anomaly arises from the one-loop diagram, 
driven by singly enhanced lepton flavor violating $\rho_{\tau\mu}$ coupling,
it can also induce rather large muon EDM, 
accessible at upcoming experiments.

\end{abstract}

\maketitle
%
\section{Introduction}\label{sec: I}

All charged-parity violation (CPV) observed in the laboratories 
so far are in the quark sector.
The Cabibbo-Kobayashi-Maskawa (CKM) 
matrix~\cite{Cabibbo:1963yz,Kobayashi:1973fv}
that govern the weak interactions of the Standard Model (SM)
has been impressively successful in accounting for all such effects.
But, as is known, the KM phase does not
provide enough CPV to satisfy 
the Sakharov condition~\cite{Sakharov:1967dj} for
the disappearance of antimatter in the Universe.
Various new physics (NP) models attempt to make up for
this by supplying new complex phases to hopefully explain 
the baryon asymmetry of the Universe (BAU).

The general two Higgs doublet model
(g2HDM)~\cite{Branco:2011iw, Davidson:2005cw}, also known as 
2HDM Type-III~\cite{Hou:1991un}, is attractive in this context.
In g2HDM, one does not introduce a $Z_2$ symmetry to
enforce Natural Flavor Conservation~\cite{Glashow:1976nt}. 
Thus, the second Higgs doublet brings in extra Yukawa coupling matrices, 
which are in general {\it complex}. The new phases, 
especially those associated with extra top quark couplings, 
$\rtt$ and $\r_{tc}$, which are likely the stronger in strength, 
can drive electroweak baryogengesis  (EWBG)
and explain BAU~\cite{Fuyuto:2017ewj,Fuyuto:2019svr}.
Complex extra Yukawa couplings can also contribute to
CPV observables accessible in the laboratory,
such as charged lepton electric dipole moment ($\ell$EDM), 
which we denote as $d_\ell$.
%

Lepton EDMs are powerful probes of NP CPV.
The SM predictions for $\ell$EDM are extremely tiny
(e.g. $|d_e|\sim 10^{-38} \; e$
\,cm~\cite{Pospelov:2005pr,Pospelov:2013sca,Smith:2017dtz})
because the sole complex phase, the KM phase, 
survives only at the four-loop level.
With essentially no SM background, detection of $\ell$EDM
will establish the presence of NP of CPV nature.

Recently, there has been very good progress in the search for $e$EDM.
The ACME experiment exploits the large internal electric field 
of ThO molecules to measure $e$EDM. 
Already in 2013, ACME provided the best 
upper limit on $e$EDM,
$|d_e|< 8.7\times 10^{-28}$~$e$~cm~\cite{ACME:2013pal}.
Impressively, the bound was further improved to~\cite{ACME:2018yjb} 
%
\begin{align}
	|d_e| < 1.1 \times 10^{-29}\;e\,{\rm cm},~~\;({\rm ACME\,2018})
\label{eq: ACME2018}
\end{align}
which is the best upper limit on any EDM.
The bound of Eq.~\eqref{eq: ACME2018} puts stringent constraint
on any BSM model that seek to address the BAU problem.

For g2HDM, consistency with  Eq.~\eqref{eq: ACME2018} has 
helped shed light on the structure of extra Yukawa couplings. 
As noted in Ref.~\cite{Fuyuto:2019svr},
one can accommodate EWBG in g2HDM
with flavor conserving coupling $|\rtt| \gtrsim 0.1$
(see Ref.~\cite{Fuyuto:2017ewj} for flavor violating coupling
$\r_{tc}$ which is immune to $e$EDM constraint),
but parameter space with single coupling scenario 
runs into trouble with the ACME\;2018 bound.
This can be avoided by introducing extra electron Yukawa coupling 
$\r_{ee}$, which helps salvage parameter space favorable for EWBG. 
Interestingly, the organization of these couplings follow the pattern~\cite{Fuyuto:2019svr},
\begin{align}
	|\r_{ee}/\rtt| \propto \l_e/\l_t,
\label{eq: top-e}
\end{align}
suggesting that the extra $\rho$ couplings of g2HDM have 
similar hierarchical structure as Yukawa couplings in SM.
Eq.~\eqref{eq: top-e} provides not only an elegant way of 
complying with the ACME 2018 bound, it connects 
EWBG with tiny CPV effects in the laboratories, 
such as $e$EDM.

Seeing that $e$EDM is highly sensitive to extra Yukawa couplings, 
in this paper we wish to explore the detailed contributions of
extra Yukawa couplings to $\mu$ and $\tau$ EDM.
Such an investigation is well-timed for the muon,
where the Fermilab Muon g-2 experiment~\cite{Muong-2:2021ojo}
has affirmed recently the significant deviation from SM  expectation
in a related observable --- the anomalous magnetic moment of the muon.
Any NP in muon $g-2$ certainly can have important implications 
for $\mu$EDM, as we will show later. 

The current bound on $\mu$EDM is by the Muon g-2 experiment
at Brookhaven National Laboratory (BNL)~\cite{Muong-2:2008ebm},
\begin{align}
	|d_\mu| <  1.8\times 10^{-19}\;e\,{\rm cm},~~\;({\rm BNL\, 2008})
\label{eq: dmu-exp}
\end{align}
while an indirect limit, extracted from $e$EDM bound of ACME\,2018,
is claimed~\cite{Ema:2021jds} at $\sim 2 \times 10^{-20}\;e\,{\rm cm}$.
In the coming years, several experiments will search for $\mu$EDM
and improve Eq.~(\ref{eq: dmu-exp}).
The Fermilab Muon g-2~\cite{Chislett:2016jau} and  
the J-PARC muon g-2/EDM~\cite{Abe:2019thb} experiments 
project sensitivities at $\sim 10^{-21}\;e$\,cm.
Another experiment being planned at PSI~\cite{Adelmann:2021udj} 
aims at improving by at least another order of magnitude, 
eventually reaching a sensitivity of $6\times 10^{-23}\;e$\,cm.
Thus, prospects for $\mu$EDM in the coming decade appears promising.

{
\begin{table*}[t!]
\begin{center}
\begin{tabular}{|c|l|l|}
\hline
\ lepton ($\ell$)   \ &
 \quad\quad\quad \ Current bound ($e$\,cm) & 
 \quad\quad \ \ Future sensitivity ($e$\,cm)   \\	
\hline\hline
electron 	&  
\hskip 1cm $1.1\times 10^{-29}$ (ACME~\cite{ACME:2018yjb}) 	&  
\ \hskip 2cm {-volatile-} 	\\
 & & \\
muon 	&  
\hskip 1cm $1.8\times 10^{-19}$ (BNL~\cite{Muong-2:2008ebm}) \ 	& 
\ $\sim 10^{-21}$ (FNAL\;\cite{Chislett:2016jau}, J-PARC~\cite{Abe:2019thb})  	\\
& \hskip 0.91cm {$\sim 2 \times 10^{-20}$ (Pospelov~\cite{Ema:2021jds})} & 
\ $\sim 6 \times 10^{-23}$\,-- \ \, (PSI~\cite{Adelmann:2021udj}) \\
 & & \\
tau & 
\ \ $\operatorname{Re} d_\tau = (-0.62 \pm 0.63) \times 10^{-17}$ (Belle~\cite{Belle:2021ybo}) \ & 

\ $\sim 10^{-18}{\rm -}10^{-19}$ (Belle~II~\cite{Belle-II:2018jsg})  \\
& & 
\ {$\sim 0.7 \times 10^{-19}$} \;\, (Bernreuther \cite{Bernreuther:2021elu}) \  \\
 & 
\ \ $\operatorname{Im} d_\tau = (-0.40 \pm 0.32) \times 10^{-17}$ (Belle~\cite{Belle:2021ybo}) \ & 

\ $\sim 10^{-18}{\rm -}10^{-19}$ (Belle~II~\cite{Belle-II:2018jsg})  \\

& & 
\ {$\sim 0.4 \times 10^{-19}$} \;\, (Bernreuther \cite{Bernreuther:2021elu}) \  \\
\hline
\hline
\end{tabular}
\caption{Present bounds and 
future sensitivities for EDM of charged leptons, $|d_\ell|$.}
\end{center}
\label{tab: dlep}
\end{table*}
}

Compared to $e$EDM and $\mu$EDM, 
$\tau$EDM measurement is
hampered by the short $\tau$ lifetime.
The present limits are derived from
spin-momentum correlations of final state decay products in 
$e^+e^-  \to \g^\ast \to \tau^+\tau^-$,
where $\tau$EDM enters 
the matrix element of the $\g\tau\tau$ vertex.
An old limit of the Belle experiment~\cite{Belle:2002nla} 
with $29.5$\;fb$^{-1}$ was very recently
updated~\cite{Belle:2021ybo} to full dataset of 833\;fb$^{-1}$.
These limits on the real and imaginary parts of $d_\tau$, 
at $\sim 10^{-17}\;e$\,cm, can be read off Table\;\ref{tab: dlep}.
An indirect bound extracted from $e$EDM~\cite{Ema:2021jds}
seems much better, but real and imaginary parts were not separately given.
%
Belle\;II~\cite{Belle-II:2018jsg} should
improve the bound 
by more than an order of magnitude.
Furthermore, some improvement 
is possible by studying 
``optimal'' observables~\cite{Bernreuther:2021elu}.

If one assumes that the NP contribution scales as $m_\ell$,
such as models with~\cite{DAmbrosio:2002vsn} minimal flavor violation 
(MFV), one expects $|d_\mu/d_e|\simeq m_\mu/m_e$.
Taking the current limit on $e$EDM of Eq.~\eqref{eq: ACME2018},
$d_\mu$ is expected at only $\sim 2 \times 10^{-27}\;e$\,cm.
A glance at Table~\ref{tab: dlep} shows that future sensitivities 
fall short by more than four orders of magnitude, 
and it is even farther away for $d_\tau$.
But this $m_\ell$ scaling may not hold in NP models 
where the underling flavor dynamics is quite different~\cite{Hiller:2010ib,Crivellin:2018qmi}, 
and the situation need not be as {\it pessimistic}. 
The g2HDM provides a simple example
where extra Yukawa couplings can 
give much larger $d_\mu$ and $d_\tau$.
We will show that, in a scenario favored by the muon $g-2$ anomaly,
very large $d_\mu$ that can be probed in the coming years is possible.

This paper is organized as follows.
In the next section, we give the extra Yukawa interactions in g2HDM 
and our working assumptions for their strength.
We then discuss $\mu$EDM in Sec.~\ref{sec: II} 
and present results for the one- and two-loop contributions,
while in Sec.~\ref{sec: III} we analyze $\tau$EDM in a similar fashion.
In Sec.~\ref{sec: IV}, we discuss extra {Yukawa} couplings that
violate our working assumptions, as possibly hinted by muon $g-2$, 
and discuss implications for EDM of all three charged leptons.
Finally, we offer our conclusion in Sec.~\ref{sec: V}.

\section{Preliminaries}
\label{sec: prelim}

Before we proceed to the specifics,
let us lay out the framework of the model.
The Yukawa interactions in g2HDM 
are given by~\cite{Davidson:2005cw, Hou:2017hiw},
\begin{align}
\mathcal{L} = 
 - & \frac{1}{\sqrt{2}} \sum_{f = u, d, \ell} 
 \bar f_{i} \Big[\big(\lambda^f_i \delta_{ij} s_\gamma + \rho^f_{ij} c_\gamma\big) h \nn\\
 & + \big(\lambda^f_i \delta_{ij} c_\gamma - \rho^f_{ij} s_\gamma\big)H
    - i\,{\rm sgn}(Q_f) \rho^f_{ij} A\Big]  R\, f_{j} \nn\\
 & - \bar{u}_i\left[(V\rho^d)_{ij} R-(\rho^{u\dagger}V)_{ij} L\right]d_j H^+ \nn\\
 &- \bar{\nu}_i\rho^\ell_{ij} R \, \ell_j H^+
 +{h.c.},
\label{eq: Lag}
\end{align}
where $L\,(R)=(1 \mp\gamma_5)/2$ are projection operators,
$Q_f$ is the charge of fermion $f$ in units of $e$,
and $V$ is the CKM matrix. 
The shorthand $\cg\equiv \cos \gamma$
and $s_\gamma \equiv \sin \gamma$ represent
mixing between two CP even scalars, $h$ and $H$,
which is known to be very small.

Although $\r_{ij}$ couplings are generic and do not possess
information about SM Yukawa couplings $\l_i$,
it is intriguing that data suggest the $\r$ matrices 
imitate closely the hierarchical structure of
SM Yukawa matrices (see Eq.~(\ref{eq: top-e})), 
as already mentioned in the Introduction.
The $\Delta F=2$ processes such as
$B_{s (d)}-\overline B_{s (d)}$ and $K^0-\bar K^0$ mixing 
constrain off-diagonal entries of down-type $\r^d$ to very small values~\cite{Crivellin:2013wna},
which indicates that $\r^d$ is nearly diagonal.
On the other hand, the extra top Yukawa coupling $\r_{tt}$,
important for our study,
is allowed to be significant in strength~\cite{Hou:2020chc}.
For the lepton sector, the $\r_{\ell\ell^\prime}$ coupling
related to third and second generation are tightly
constrained from LFV decays $h\to\tau\mu$ and $\tau\to\mu\g$, 
but with $\rtt\sim 1$, values of $\r_{\tau\mu}=\r_{\mu\tau}\lesssim\o(\l_\tau)$ 
are easily allowed even for $\cg \sim 0.1$~\cite{Hou:2020tgl}.

With these arguments in mind,
we assume the following strengths for 
$\r_{ij}$ related to $u$- and $\ell$-type matrices~\cite{Hou:2020itz},
\begin{equation}\label{eq: rho}
	\begin{aligned}
		\r_{ii}\lesssim \o(\l_i);\;\; \r_{1i}\lesssim\o(\l_1);
		\;\; \r_{3j}\lesssim\o(\l_3)\; (j\ne 1),
	\end{aligned}
\end{equation}
where the first two conditions can also be applied to
elements of the $\r^d$ matrix.
Down-type couplings, however, do not play
much role in our discussion, so
Eq.~\eqref{eq: rho} will be our working assumption while  
estimating the typical size of $\ell$EDM in g2HDM.
But we note that deviation from Eq.~\eqref{eq: rho} is possible.
For example, NP in muon $g-2$ may call for large deviation from 
the last condition in Eq.~\eqref{eq: rho} for interpretation 
in 2HDM~\cite{Hou:2021sfl}.
We will discuss in Sec.~\ref{sec: IV} 
the implications of such a scenario for $\ell$EDM and assess
the impact on our results.

{
It is worth pointing out that with $\cg\lesssim 0.1$
our assumptions $\rho_{\mu \mu}\lesssim\o(\l_\mu)$
and $\rho_{\tau\tau}\lesssim\o(\l_\tau)$
are easily compatible with the latest data
on di-muon and di-tau decays of the $h$ boson, respectively.
{These processes provide access to the strength and phase
of extra Yukawa couplings}.
Both ATLAS~\cite{ATLAS:2020fzp} and CMS~\cite{CMS:2020xwi}
have reported the $h\to \mu\mu$ branching fraction relative to SM at
$1.2 \pm 0.6$ and $1.19^{+0.40+0.15}_{-0.39-0.14}$, respectively.
The $h\mu\mu$ coupling in Eq.~\eqref{eq: Lag} induces the following
modification in g2HDM,
\begin{align}
	\frac{{\cal B}(h \to \mu\mu)}{{\cal B}(h \to \mu\mu)_{\rm SM}} 
	= \left| s_\gamma + \cg\operatorname{Re}\frac{\rmm}{\l_\mu} \right|^2
	+  \left|\cg\operatorname{Im}\frac{\rmm}{\l_\mu} \right|^2.
\label{eq: rat-hmumu}
\end{align} 
For $\cg=0.1$ and
$\operatorname{Re}\rmm = \operatorname{Im}\rmm=\l_{\mu}/\sqrt{2}$,
Eq.~\eqref{eq: rat-hmumu} gives $\simeq 1.14$,
consistent with data.
}

{
The $h\to\tau\tau$ process provides an avenue
for the study of CP properties of the $h$ boson at the LHC 
(see e.g. Refs.~\cite{DellAquila:1988bko, Harnik:2013aja}).
Very recently, CMS~\cite{CMS:2020rpr} has carried out
a study of the CP structure of the $h\tau\tau$ coupling with $137$\;fb$^{-1}$ data,
reporting the 
{phase of the $\tau$ Yukawa coupling, $\phi_{\tau\tau}$}, at
$4^\circ \pm 17^\circ (\pm 36^\circ)$ at the $68\%\, (95\%)$ C.L.,
{with rather large errors.}
In the notation of Ref.~\cite{CMS:2020rpr},
the $h\tau\tau$ coupling in Eq.~\eqref{eq: Lag} gives,
\begin{align}
	\tan \phi_{\tau\tau}
	= \frac{\cg \operatorname{Im}\rtata}{s_\g \l_\tau
	+ \cg \operatorname{Re}\rtata}.
\end{align}
For $\cg=0.1$ with
$\operatorname{Re}\rtata=\operatorname{Im}\rtata=\l_\tau/\sqrt{2}$,
we find $\phi_{\tau\tau}\simeq 3.8^\circ$, which is compatible with data.
A more precise measurement of $\phi_{\tau\tau}$ in the future could
constrain this extra $\tau$ Yukawa coupling.  
}

\begin{figure}[b]
\center
\includegraphics[width=.23\textwidth]{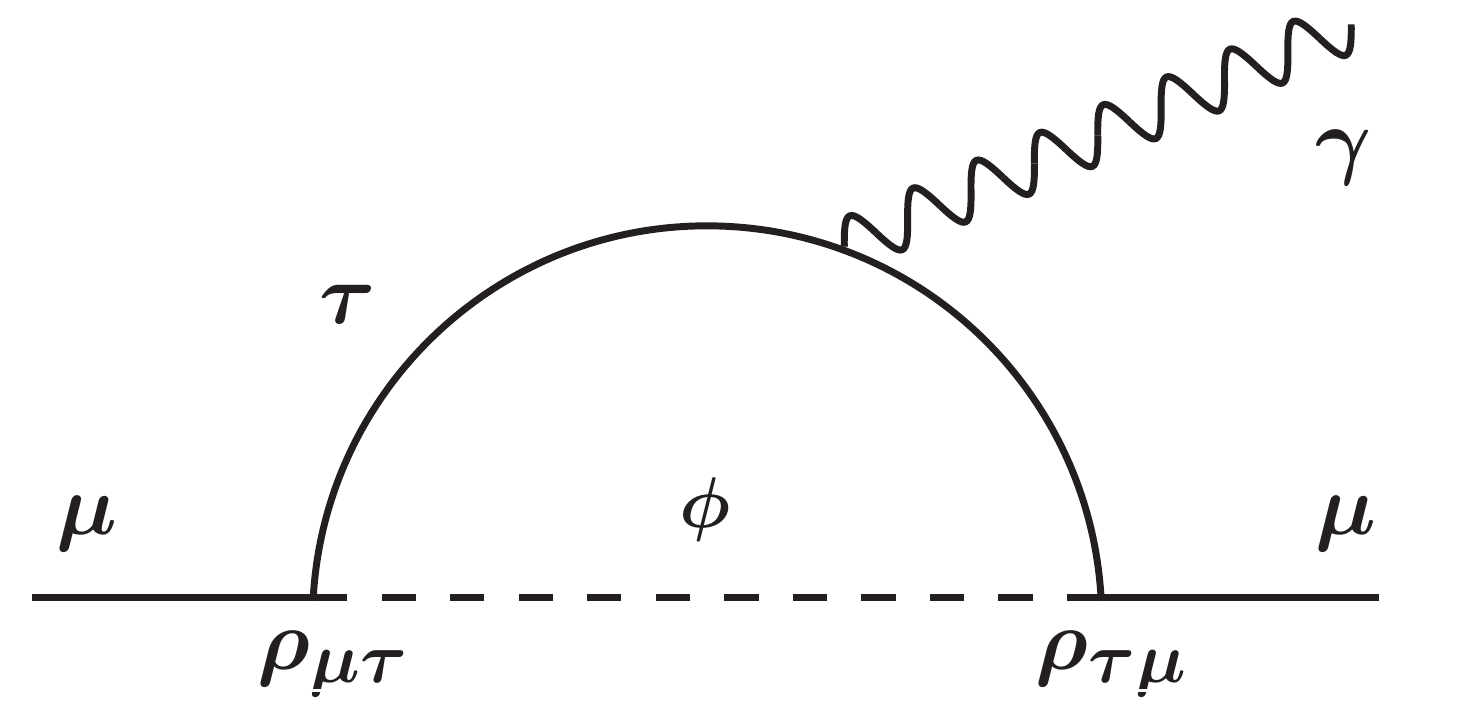}
\caption{One-loop digram for $\mu$EDM. }
\label{fig: 1loop}
\end{figure}

\section{\boldmath$\mu$EDM}
\label{sec: II}

The EDM, $d_\ell$, of lepton $\ell$ is defined through the
effective Lagrangian,
\begin{equation}\label{eq: eff-L}
	\begin{aligned}
\mathcal{L}_{\mathrm{eff}}^{\rm EDM} =
 -\frac{i}{2}d_\ell\, \bar{\ell}\,
 \sigma_{\alpha\beta}\g_5  \ell \,F^{\alpha\beta},
\end{aligned}
\end{equation}
where $\sigma_{\alpha\beta} = i[\g_\alpha, \g_\beta]/2$,
and $F^{\alpha\beta}$ is the electromagnetic field
strength tensor.

In g2HDM, the first finite contribution to $\ell$EDM appears at one-loop.
A sample one-loop diagram for $\mu$EDM 
in g2HDM is given in Fig.~\ref{fig: 1loop}.
The dipole operator in Eq.~\eqref{eq: eff-L} is chirality violating,
so an additional mass insertion
is required on the fermion line to obtain the correct
chiral structure. 
This means that analogous diagrams with 
lighter leptons in the loop are chirally suppressed,
and we neglect them.  
Fig.~\ref{fig: 1loop} gives~\cite{Omura:2015xcg},
\begin{align}
 d_\mu|_{\rm 1-loop}
	\simeq & -e \frac{ m_\tau}{32\pi^2}
		\operatorname{Im}(\rho_{\tau\mu}\rho_{\mu\tau})
		\left[\cg^2\, \frac{\log x_{h\tau}-3/2}{m_h^2}\right. \nonumber \\
	& +\left. s^2_\g\, \frac{\log x_{H\tau}-3/2}{m_H^2}-\frac{\log x_{A\tau}-3/2}{m_A^2}\right],
\label{eq: 1loop}
\end{align}
where
$x_{ij}=m_i^2/m_j^2$.

We note that the contribution of the $h$ boson is negligible 
in Eq.~\eqref{eq: 1loop} because $\cg^2\ll1$,
while the contributions of CP even and odd scalars
have a relative sign between them, indicating that
in the $H$--$A$ degeneracy limit,
one-loop contributions will be highly suppressed, which is indeed the case.
For example, taking $\cg=0.1$ and $m_H=m_A=$ 300\;GeV, we get
$|d_\mu|\simeq 4 \times 10^{-22} |\operatorname{Im}(\r_{\tau\mu}\r_{\mu\tau})|\;e$\,cm.
With $|\rho_{\tau\mu}|=|\rho_{\mu\tau}|\lesssim \l_\tau$, Eq.~\eqref{eq: rho}, 
we obtain
$|d_\mu|\lesssim 4 \times 10^{-26}\;e$\,cm at one-loop, which is 
three orders of magnitude below PSI sensitivity. 
With $m_A$ and $m_H$ sufficiently apart,
the one-loop contribution increases,
but it remains outside of experimental reach.
For example, with $m_H=300$\;GeV and $m_A=500$\;GeV
and keeping other values as before,
we obtain $|d_\mu|\lesssim 7 \times 10^{-25}\;e$\,cm.
Thus, we find that our working assumption of
$\rho_{\ell\ell^\prime}$ in Eq.~\eqref{eq: rho}
renders the one-loop contribution too small. 
Next, we consider two-loop diagrams --- also known as
Barr-Zee or Bjorken-Weinberg diagrams~\cite{Barr:1990vd, Bjorken:1977vt}.

\begin{figure}[b]
\center
\includegraphics[width=.25\textwidth]{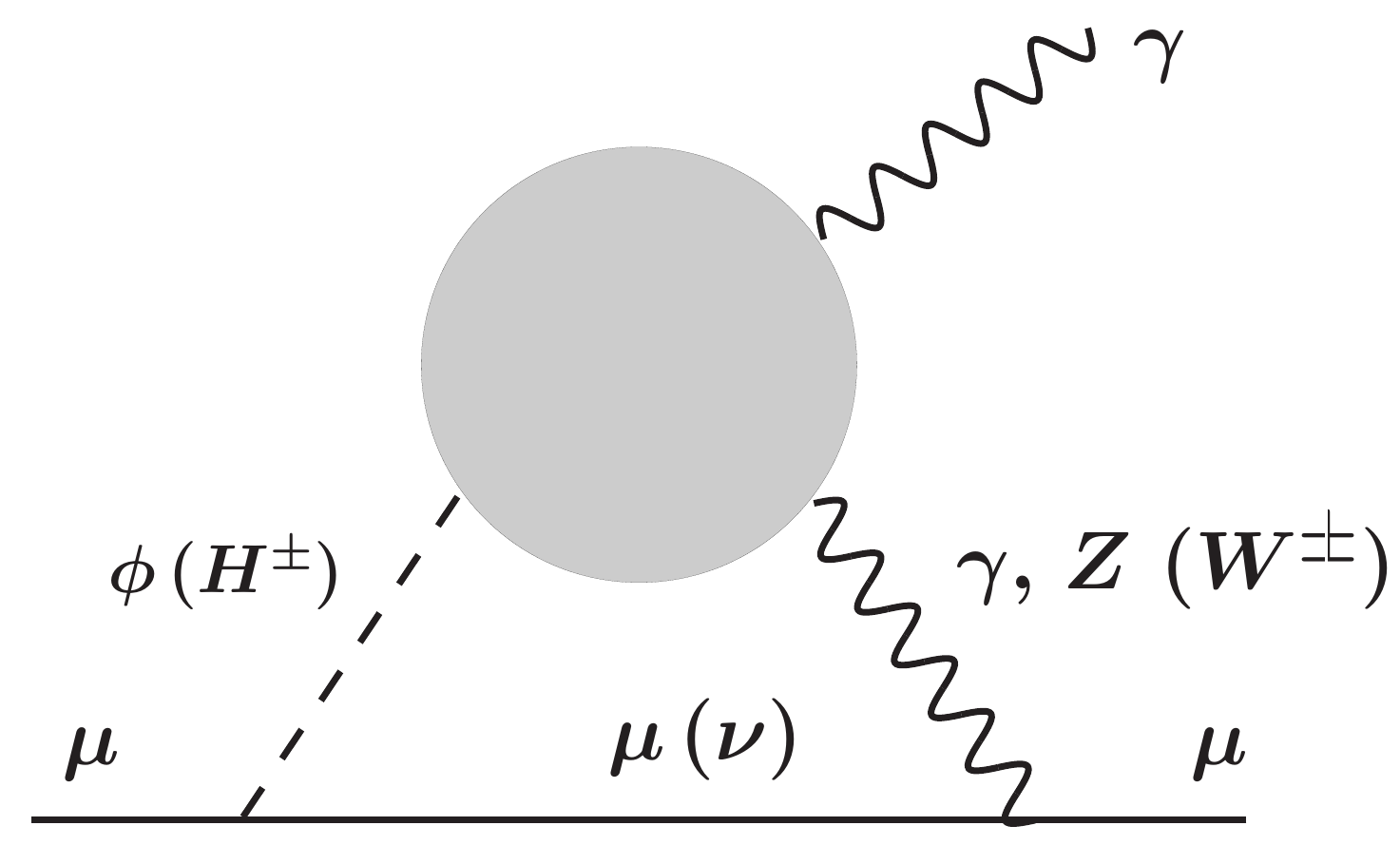}
\caption{
Typical two-loop diagram for $\mu$EDM, 
where $\phi$ ($H^\pm$) is a neutral (charged) scalar, and
the gray blob stands for fermion,
gauge, or scalar loop depending on the attached legs.}
\label{fig: 2loop}
\end{figure}

It is well known that, despite having additional loop suppression, 
certain two-loop diagrams can dominate over one-loop diagrams.
This can be understood qualitatively by noting that 
Fig~\ref{fig: 1loop} contains three chirality flips 
while only one flip in Fig.~\ref{fig: 2loop},
which can let the latter compensate for the additional
loop factor ($\sim \alpha/\pi$). 
Barr and Zee~\cite{Barr:1990vd} calculated neutral scalar contributions 
with top quark and gauge bosons in the loop, which was later 
extended~\cite{Leigh:1990kf, Chang:1990sf, Kao:1992jv, Bowser-Chao:1997kjp, Abe:2013qla}
to include other two-loop diagrams.

Following the convention used in Ref.~\cite{Fuyuto:2019svr},
we categorize two-loop contributions to $\mu$EDM into 
three parts,
\begin{equation}
	d_\mu|_{\rm 2-loop} = d_\mu^{\phi\g} + d_\mu^{\phi Z } + d_\mu^{H^+ W^+},
\label{eq: dmu}
\end{equation}
related to effective $\phi\gamma\gamma$,
$\phi\gamma Z$ and $H^+\gamma W^+$ vertices, 
respectively, as in Fig~\ref{fig: 2loop}.

For $d_\mu^{\phi\g}$ with fermions in the loop, the dominant contribution
comes from the top quark. For convenience, it is written as 
the sum of two terms~\cite{Fuyuto:2019svr},
\begin{equation}
	 (d_\mu^{\phi\g})_{t} = (d_\mu^{\phi\g})_{t}^{\rm mix}
	 						+ (d_\mu^{\phi\g})_{t}^{\rm extra}
\label{eq: dmu-top}
\end{equation}
where
%
\begin{align}
	& \left(d_\mu^{\phi\g}\right)_{t}^{\rm mix}
	 =  \frac{e\,\alpha\, s_\g\, c_\g }{6\sqrt{2}\pi^3 v}
		 \Bigl[\operatorname{Im}\r_{\mu\mu}\,\Delta f(x_{th}, x_{tH})\Bigr. \nonumber\\
	& \hskip2.28cm
 + \Bigl.\frac{m_\mu}{m_t}\operatorname{Im}\r_{tt}\,\Delta g(x_{th}, x_{tH})\Bigr],
\label{eq: dmu-mix} \\
%
%
		& \left(d_\mu^{\phi\g}\right)_{t}^{\rm extra}
		=\frac{e\, \alpha }{12\pi^3m_t} \nonumber \\
		&\hskip0.3cm\times
		\Bigl\{
		\operatorname{Im}\rho_{\mu\mu}\operatorname{Re}\rho_{tt}
		\left[ \cg^2 f(x_{th})+s_\g^2 f(x_{tH}) + g(x_{tA})
		\right]
		\Bigr. \nonumber\\ 
		&\hskip0.5cm + \Bigl.
		\operatorname{Re}\rho_{\mu\mu}\operatorname{Im}\rho_{tt}
		\left[
		 \cg^2 g(x_{th})+s_\g^2 g(x_{tH}) + f(x_{tA})
		 \right]
		\Bigr\}.
\label{eq: dmu-extra}
\end{align}
where $\Delta f(a, b) = f(a)-f(b)$,  $\Delta g(a, b) = g(a)-g(b)$,
and loop functions $f(x)$ and $g(x)$ are given in Appendix~\ref{app: fun}. 
Note that for $x\sim 1$, $f(x), g(x)\sim x$,
and for small x, $f(x), g(x)\sim x (\log x)^2/2$, which further
exemplifies why the top quark diagram is dominant.

Eq.~\eqref{eq: dmu-mix} is due to mixing of SM Yukawa couplings
with extra Yukawa couplings, and therefore has a 
linear dependence on $\operatorname{Im} \rho_{ii}$ with no dependance on $m_A$.
On the other hand, Eq.~\eqref{eq: dmu-extra} contains contributions
from extra Yukawa couplings only. 
For very small $\cg$, the ``mix'' term becomes negligible,
and the top loop contribution
$(d_\mu^{\phi\g})_{t}\simeq (d_\mu^{\phi\g})_{t}^{\rm extra}$
is governed by Eq.~\eqref{eq: dmu-extra}. 
One also notes that, for $\rmm$ and $\rtt$ both purely real or purely imaginary,
Eq.~\eqref{eq: dmu-extra} does not contribute, which yields
$(d_\mu^{\phi\g})_{t}\simeq (d_\mu^{\phi\g})_{t}^{\rm mix}$.

\begin{figure}[t]
\center
\includegraphics[width=.45\textwidth]{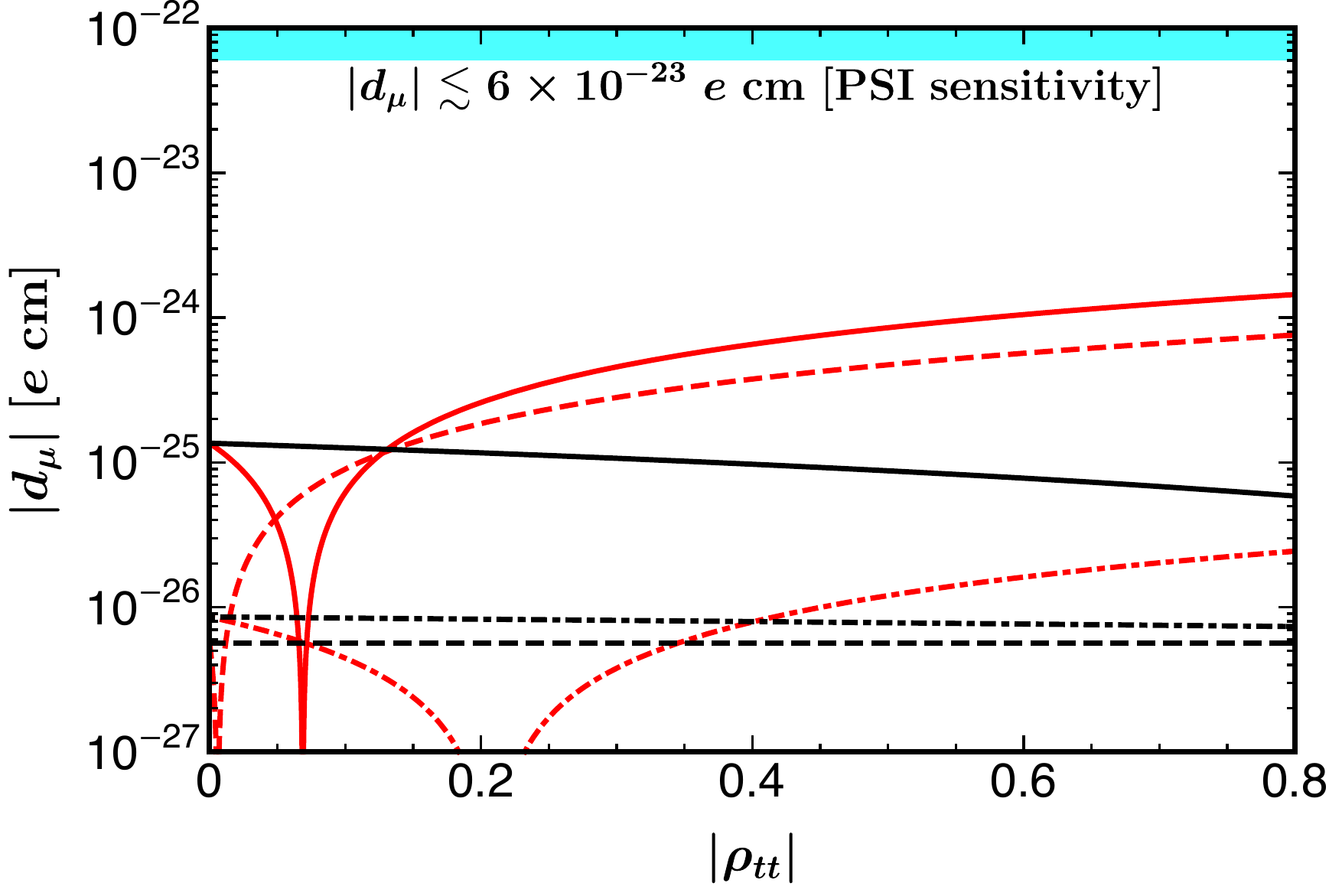}
\caption{
Different two-loop contributions to $|d_\mu|$ as function of
$\rtt$, with 
$\cg=0.1$, $m_H=m_A=m_{H^\pm}=300$ GeV, 
and $\rmm =i \l_\mu$ for illustration.
Red (black) curves correspond to $\phi_{tt}=0 \,(\pi/2)$.
Solid, dashed, and dot-dashed curves represent contributions
of $|d_\mu^{\phi\g}|$, $|d_\mu^{H^+ W^+}|$, and  $|d_\mu^{\phi Z }|$,
respectively.
}
\label{fig: muEDM}
\end{figure}

The diagram with $W^+$ in the loop induces contribution to
$d_\mu^{\phi\g}$ for finite $\cg$.
Defining the $\phi W^+ W^-$ coupling
as $i\, gm_W C_{\phi W W}$~\cite{Gunion:2002zf,Davidson:2010xv},
where $C_{hWW} = s_\g$, $C_{HWW} = c_\g$, and $C_{AWW} = 0$~,
the diagram gives,
\begin{align}
		\left(d_\mu^{\phi\g}\right)_{W}
		=-\frac{e\, \alpha s_\g c_\g }{32\sqrt{2}\pi^3 v}
		\operatorname{Im}\r_{\mu\mu} \left[ I_\g(m_h)-I_\g(m_H)\right],
\label{eq: dmu-W}
\end{align}
with loop function $I_\g(a)$ given in Appendix~\ref{app: fun}.
This diagram has sign opposite the top-loop, 
{
hence large cancellation can occur in some parameter space for finite $\cg$.
} 
The cancellation is, however, sensitive to complex phases of extra Yukawa couplings.

Adding Eqs.~\eqref{eq: dmu-top} and \eqref{eq: dmu-W} and taking
$\cg=0.1$, $m_H=m_A=300$ GeV, we illustrate $d_\mu^{\phi\g}$ 
as solid lines in Fig.~\ref{fig: muEDM} vs $\rtt$ coupling. 
We take $|\rho_{\mu\mu}|= \l_\mu $ in accordance
of Eq.~\eqref{eq: rho} and assume it to be purely imaginary.  
The red (black) lines correspond to 
purely real (imaginary) $\rtt$.
For $\phi_{tt}=0$, as $\rtt$ increases, top-loop diagram 
starts to compete with $W$-loop,
resulting in cancellation between the two, seen as the dip in the plot.
For larger values of $\rtt$, the top-loop overwhelms the $W$-loop,
thereby bringing $d_\mu^{\phi\g}$ to $\sim 10^{-24}\;e$\,cm.
For $\phi_{tt}=\pi/2$,  $(d_\mu^{\phi\g})_t$
is given by Eq.~\eqref{eq: dmu-mix} only, 
where the $\rtt$ term is suppressed further by $m_\mu$, 
hence $(d_\mu^{\phi\g})_t$ is feebly sensitive to $\rtt$ 
and never overpowers $W$-loop even for large $|\rtt| \sim 1$.

For $d_\mu^{\phi Z}$, one simply replaces the inner photon propagator
(one connecting loop with muon line) with the $Z$ propagator.
By parity, only the vector part of $Z\mu\mu$ coupling contributes.
Its smallness, together with $1/m_Z^2$ from the $Z$ propagator,
the $d_\mu^{\phi Z}$ effect is typically suppressed 
with respect to $d_\mu^{\phi \g}$.
The corresponding formulae are relegated to Appendix~\ref{app: fun}.
In Fig.~\ref{fig: muEDM}, 
$d_\mu^{\phi Z}$ are given as dot-dashed lines, with color coding as before. 
For both red and black lines,
$|d^{\phi Z}_\mu|$ always lie far lower than $|d^{\phi \g}_\mu|$,
except near cancellation regions of $|d^{\phi \g}_\mu|$.
  
On the other hand, complex extra Yukawa couplings of
charged Higgs boson in Eq.~\eqref{eq: Lag} contribute significantly
to $d_\mu^{H^+ W^+}$. 
The main effect comes from top-bottom in the loop, 
which gives~\cite{Bowser-Chao:1997kjp},
\begin{align}
 (d_\mu^{H^+ W^+})_{t/b}
	& = - \frac{3e\alpha |V_{tb}|^2\operatorname{Im}(\rtt\rmm)}
		{128\pi^3 m_t \sin^2\theta_W} \nonumber\\
	  & \hskip-0.2cm \times
		\Bigl[Q_t F_t(x_{\phi t}, x_{W t} ) + Q_b F_b(x_{\phi t}, x_{W t} )\Bigr],
\label{eq: dmu-H+}
\end{align}
where $\phi$ stands for $H^+$ only, and functions
$F_q(a, b)$ are given in Appendix~\ref{app: fun}.

Diagrams with $W^+$ and scalars in the loop
were calculated in Ref.~\cite{Abe:2013qla}.
Adapting to our case, we find,
\begin{align}
 (d_\mu^{H^+ W^+})_{W}
	& = -\, \frac{e\,\alpha s_\g \cg \operatorname{Im}\rmm}
		{128\sqrt{2}\pi^3\, v \sin^2\theta_W} \nonumber \\
	& \hskip-0.2cm \times
 \Bigl[ {\cal I}_4(m_h, m_{H^+})-{\cal I}_4(m_H, m_{H^+})\Bigr],
\label{eq: dmu-H+-W}
\end{align}
where ${\cal I}_4(a, b)$ is given in Appendix~\ref{app: fun}.

Combining Eqs.~\eqref{eq: dmu-H+} and \eqref{eq: dmu-H+-W},
we give $d_\mu^{H^+ W^+}$ in Fig.~\ref{fig: muEDM} as dashed lines,
taking the same values of scalar masses and $c_\g$ as before.
For $\phi_{tt}=0$ (red), $d_\mu^{H^+ W^+}$ is basically governed
by Eq.~\eqref{eq: dmu-H+} and slightly lower than $d_\mu^{\phi\g}$,
but significantly larger than $d_\mu^{\phi W}$. 
At $\rtt=0$, the origin, $d_\mu^{H^+ W^+}$ 
is given by Eq.~\eqref{eq: dmu-H+-W}, 
which lies between $10^{-27}$--$10^{-26}\; e$\,cm. 
For extremely tiny $\rtt$, the effects from Eqs.~\eqref{eq: dmu-H+}
and~\eqref{eq: dmu-H+-W} tend to cancel.
This behavior is similar to the case of $d_\mu^{\phi \g}$ discussed previously.
For $\phi_{tt}=\pi/2$ (black), there is no dependence on $\rtt$
and the total contribution is given by Eq.~\eqref{eq: dmu-H+-W}, 
which is represented as a horizontal black dashed line.

We see from Fig.~\ref{fig: muEDM} that, 
our working assumption of Eq.~\eqref{eq: rho} suggests 
$\mu$EDM in g2HDM can be enhanced to $\sim 10^{-24}\;e$\,cm. 
This, however, lies out of reach of muon g-2 experiments 
at FNAL and J-PARC by three orders of magnitude. 
Even with improved sensitivity at PSI, shown as
cyan band in Fig.~\ref{fig: muEDM},
it cannot probe extra Yukawa couplings 
with strength as given in Eq.~\eqref{eq: rho}.

Some comments are in order. There are additional subdominant diagrams related to 
charged Higgs loops which depend on trilinear Higgs couplings.  
The expressions of these are given in Appendix~\ref{app: fun}.
We ignore them since our scope is to cover effects from 
Yukawa interactions of
Eq.~\eqref{eq: Lag}.
These contribution can be important in models with
CPV scalar potentials, or 2HDM models with softly broken $Z_2$ symmetry, 
where large contributions such as in Eq.~\eqref{eq: dmu-H+}
are negligible (see e.g. Refs.~\cite{Abe:2013qla,
Inoue:2014nva,Chun:2019oix,Altmannshofer:2020shb, Egana-Ugrinovic:2018fpy}).
For the former case, we should mention that there are additional
so-called ``kite diagrams", calculated recently in
Ref.~\cite{Altmannshofer:2020shb},
which are required to obtain a gauge-invariant result.

\section{$\tau$EDM}\label{sec: III}
The diagrams contributing to $\tau$EDM is similar
to $\mu$EDM, except a more significant
coupling $\rtata\sim \l_\tau$ is involved, hence one would
expect further enhancement.

One obtains the one-loop effect for $\tau$EDM 
by replacing $``\mu" \to ``\tau"$ in Fig.~\ref{fig: 1loop}.
Diagrams with internal $\mu$ and $e$ are 
again neglected because of chiral suppression. 
Since the flavor conserving $\phi\tau\tau$ vertex is involved, 
the SM Yukawa $\l_\tau$ also contributes via interference with $\rho_{\tau\tau}$.
Using the results of Ref.~\cite{Crivellin:2013wna},
the one-loop contribution is
\begin{align}
 d_\tau & \simeq -\frac{e\, m_\tau}{32\pi^2} \nonumber \\
 & \,  \times
		\left\{\left[s_{2\g}\l_\tau \operatorname{Im}\rtata
                 + \cg^2\operatorname{Im}\rtata^2\right] 
                    \frac{\log x_{h\tau}-3/2}{m_h^2}\right. \nonumber\\
 & \,  + \left[-s_{2\g}\l_\tau \operatorname{Im}\rtata
 +s_\g^2\operatorname{Im}\rtata^2\right]\frac{\log x_{H\tau}-3/2}{m_H^2} \nonumber\\
 & \hskip3.22cm  -\operatorname{Im}\rtata^2\left.\frac{\log x_{A\tau}-3/2}{m_A^2}\right\}.
 \label{eq: 1loop-tauEDM}
\end{align}
In the small $\cg$ limit, $\l_\tau$ effects decouple.
Taking $\cg=0.1$, $m_H=m_A=300$ GeV, and
{$\operatorname{Re}\rtata=\operatorname{Im}\rtata \le \l_\tau/\sqrt{2}$}
for $\tau$ coupling following Eq.~\eqref{eq: rho},
we find 
{$|d_\tau|\lesssim 6 \times 10^{-25}\;e$\,cm,} which is about
five orders below Belle\;II projections.  
For $m_H=300$ GeV and $m_A=500$ GeV, 
the one-loop contribution increases to 
{$|d_\tau|\lesssim 1.3 \times 10^{-24}\;e$\,cm,} 
but remains out of Belle\;II reach.

The discussion of two-loop contribution to $d_\tau$ is analogous
to $d_\mu$ in Sec.~\ref{sec: II}.
Therefore, similar to Eq.~\eqref{eq: dmu}, we write for $d_\tau$ at two-loop,
\begin{equation}
	d_\tau|_{\rm 2-loop} = d_\tau^{\phi\g} + d_\tau^{\phi Z } + d_\tau^{H^+ W^+},
\label{eq: dtau}
\end{equation}
where expressions for each term can be obtained from Eqs.~\eqref{eq: dmu-top} 
and \eqref{eq: dmu-W}-\eqref{eq: dmu-H+-W}
by change of label $``\mu"\to ``\tau"$.

\begin{figure}[t]
\center
\includegraphics[width=.453\textwidth]{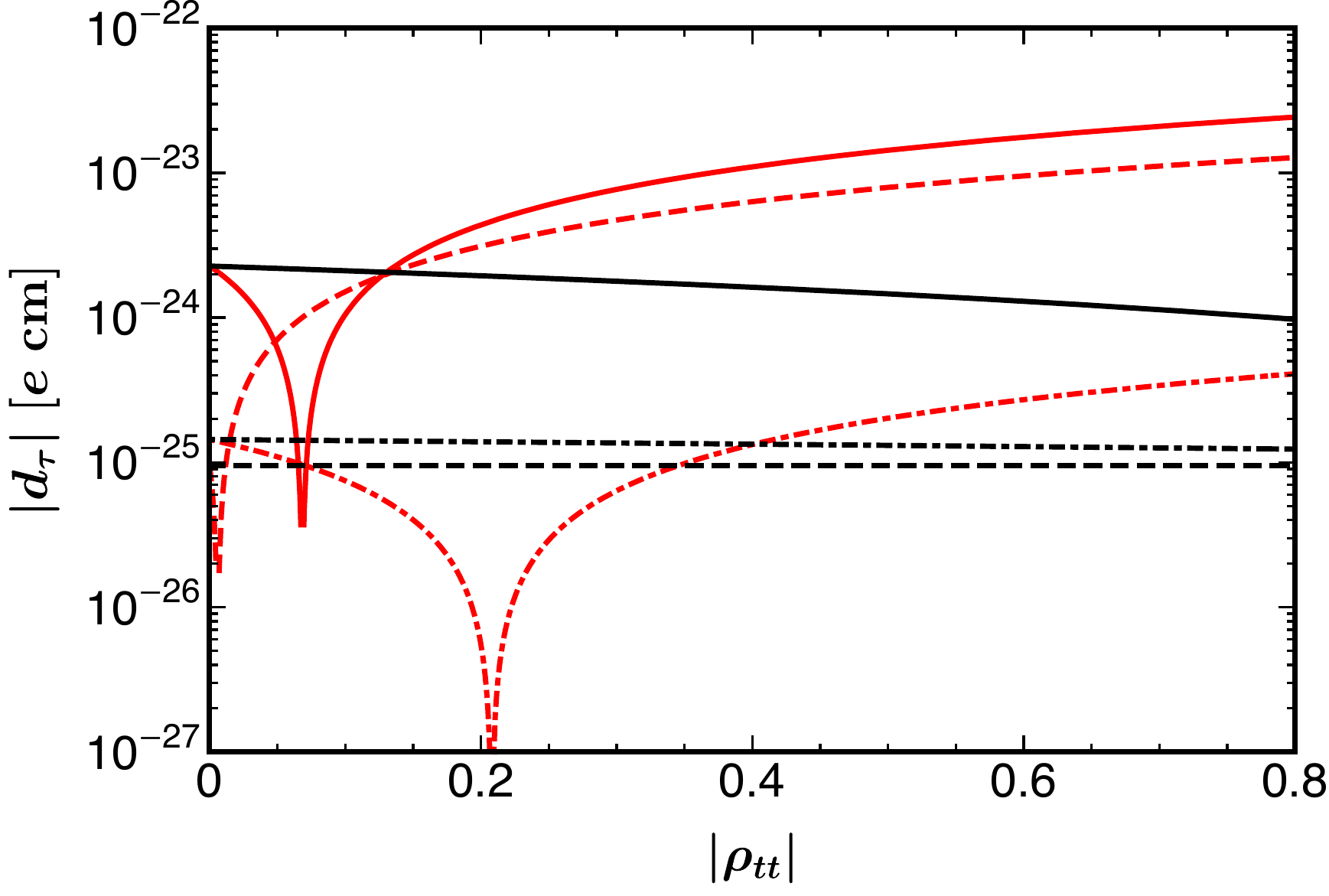}
\caption{
Different two-loop contributions to $|d_\tau|$ as function of 
 $\rtt$, taking $\rtata=i \l_\tau$.
 Color coding is the same as in Fig.~\ref{fig: muEDM}.  }
\label{fig: tauEDM}
\end{figure}

In Fig.~\ref{fig: tauEDM} we give results for each term in
Eq.~\eqref{eq: dtau} as functions of $\rtt$.
The same benchmarks are used as in previous section:
$m_H=m_A=m_{H^\pm}=$ 300~GeV, $\cg=0.1$, and we take $\rtata=i \l_\tau$.
Red (black) lines represent $\phi_{tt} = 0\, (\pi/2)$.
The characteristics of each type of contribution
are the same as the $\mu$EDM case in Sec.~\ref{sec: II}, 
except that each contribution 
is now larger by $\r_{\tau\tau}/\r_{\mu\mu}$.
From Fig.~\ref{fig: tauEDM} and for our benchmarks,
$d_\tau$ can reach $\sim 10^{-23}\;e$\,cm,
but again falls short of Belle\;II sensitivity.

%
\begin{figure*}[t]
\center
\includegraphics[width=.375\textwidth]{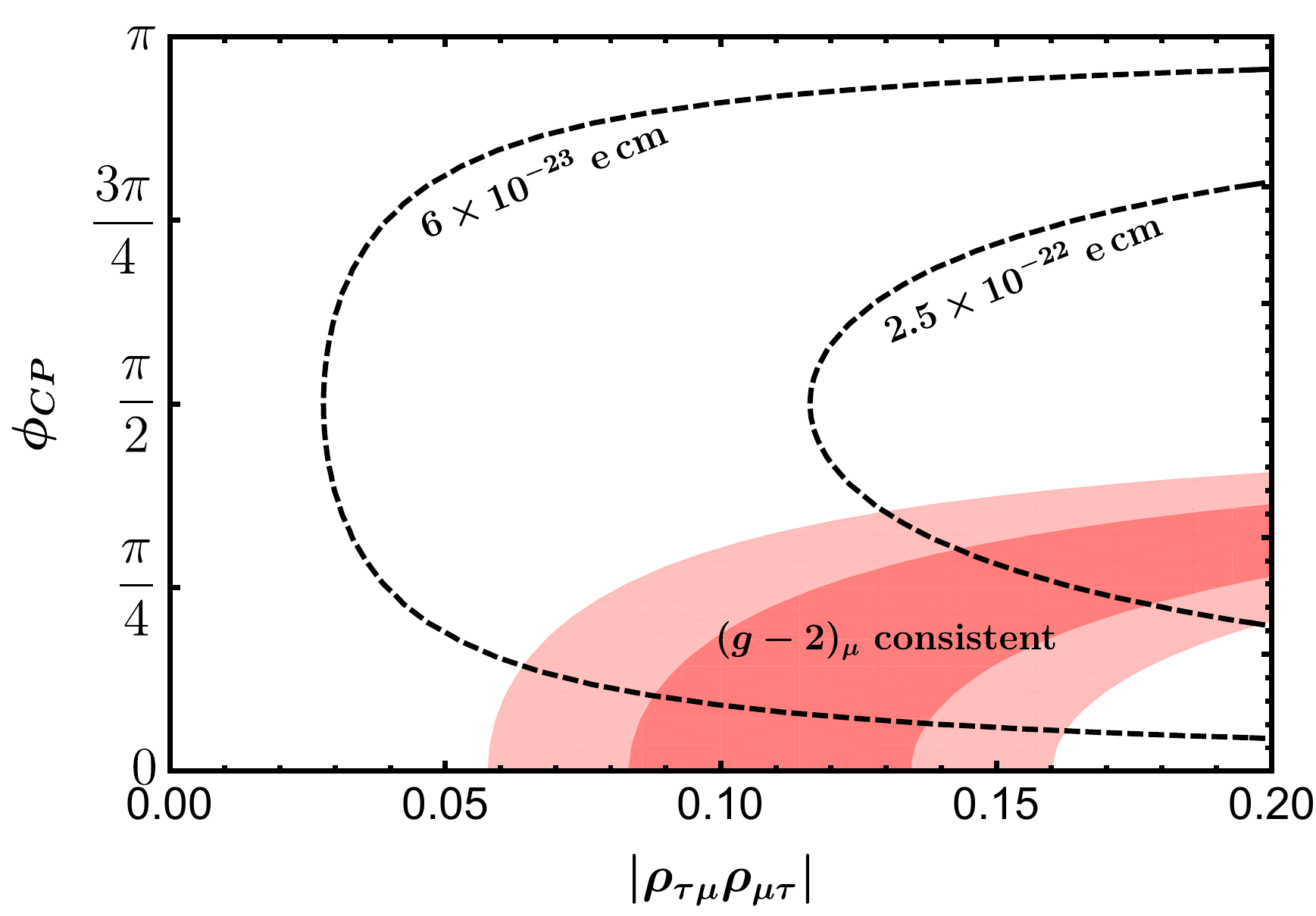}
~\includegraphics[width=.375\textwidth]{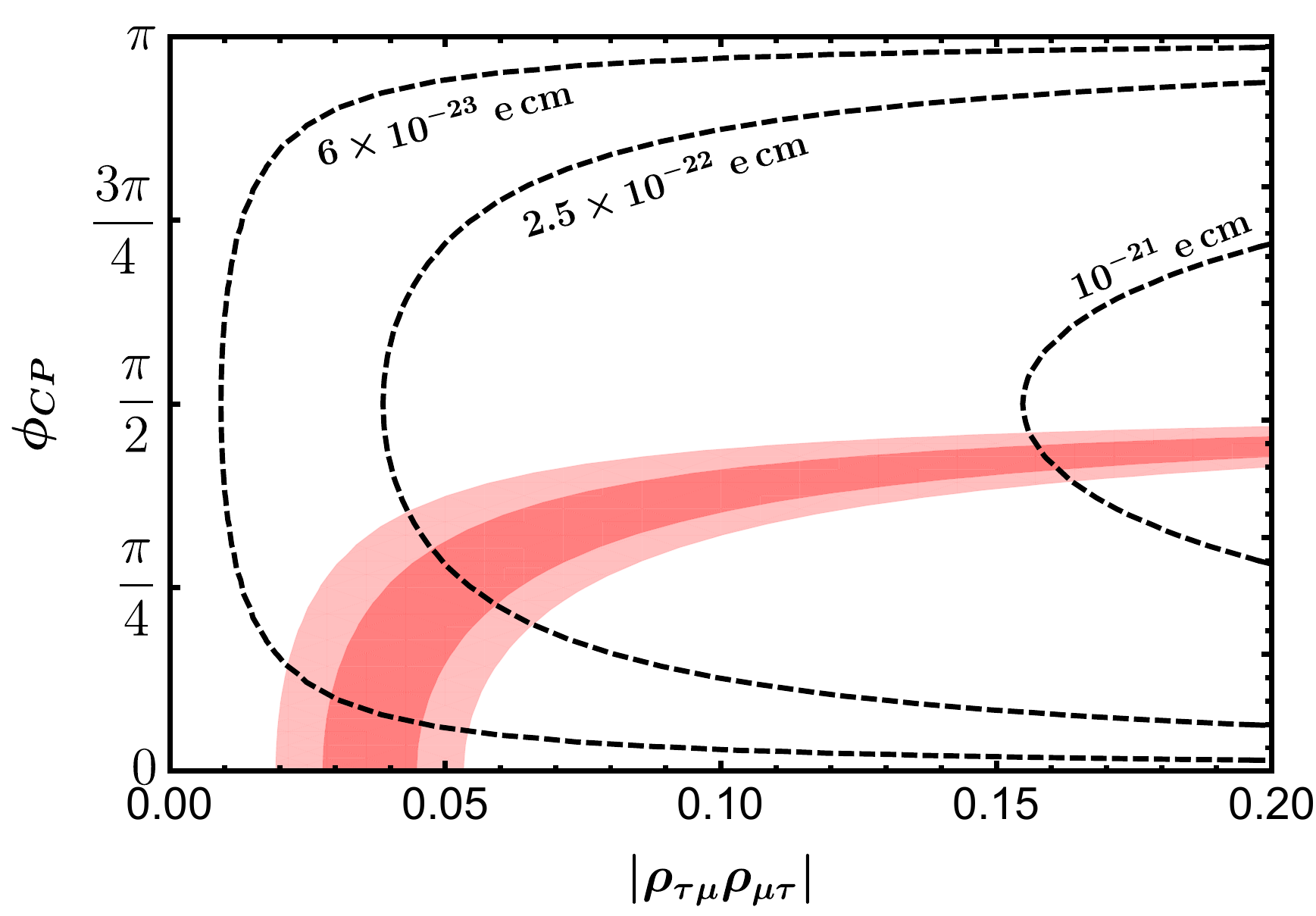}
\caption{
Correlation between muon $g-2$ and $\mu$EDM.
Pink (light pink) region shows parameter space that explain
Eq.~\eqref{eq: del_a_mu} at $1\sigma\,(2\sigma)$, 
while dashed lines are $d_\mu$ contours. 
The left (right) plot is for $\cg=0$, $m_H=$ 300~GeV
and $m_A=340$ (500)~GeV. All other $\rho_{ij}$ are set to zero.  }
\label{fig: amu-dmu}
\end{figure*}

\section{Discussion}\label{sec: IV}

With our working assumption for $\rho_{ij}$ given in Eq.~\eqref{eq: rho}
and sub-TeV exotic scalars, we find that $\mu$ and $\tau$ EDM 
can be enhanced in g2HDM, but are not large enough to be probed in the coming years. 
However, 
 recent progress in muon $g-2$ may suggest
good prospects for $\mu$EDM.

The Fermilab Muon g-2 experiment~\cite{Muong-2:2021ojo}
reported recently their first measurement of the muon anomalous 
magnetic moment, $a_\mu\equiv (g-2)_\mu$, confirming 
the old BNL result~\cite{Muong-2:2006rrc}.
The combined value of
$ a_\mu^{\rm Exp} = 116 592 061(41) \times 10^{-11}$
disagrees with the SM prediction,
$a_\mu^{\rm SM} = 116 591 810(43) \times 10^{-11}$~\cite{Aoyama:2020ynm},
by more than $4\sigma$,
\begin{align}
a_\mu^{\rm Exp} - a_\mu^{\rm SM} =  (251 \pm 59) \times 10^{-11}.
\label{eq: del_a_mu}
\end{align}

One of the simplest
(and easiest, thanks to the chiral enhancement factor
 of $m_\tau/m_\mu\simeq 17$) 
NP explanation of Eq.~\eqref{eq: del_a_mu} is provided by g2HDM 
via the same diagram as in Fig.~\ref{fig: 1loop}.
But it would require significant departure from Eq.~\eqref{eq: rho} for some couplings. 
As shown in Ref.~\cite{Hou:2021sfl}, 
to explain  Eq.~\eqref{eq: del_a_mu} at $1\sigma$, one needs
$|\r_{\tau\mu}|=|\r_{\mu\tau}|\simeq \o (20) \l_\tau$,
with $m_H$ and $m_A$ sub-TeV but nondegenerate.
%

Several implications, relevant for the present paper, follow. 
First, one has to adopt~\cite{Hou:2021sfl} near-perfect alignment, i.e. $\cg\to 0$.
Though odd, it is needed to satisfy the CMS limit of 
${\cal B}(h\to\tau\mu) < 0.15\%$~\cite{CMS:2021rsq},
which constrains $|\r_{\tau\mu}\cg|<0.1\l_\tau$ for $\r_{\tau\mu}=\r_{\mu\tau}$.
This bound implies $\cg< 0.005$ for $1\sigma$ solution of muon $g-2$.
Another important implication is a rather small $\rtt$.
The decay $\tau\to\mu\gamma$ arises from Fig.~\ref{fig: 2loop} 
with initial $\mu$ replaced by $\tau$,
which constrains the product $\r_{\tau\mu}\rtt$. 
To comply with the Belle bound of
${\cal B}(\tau\to\mu\gamma) < 4.2 \times 10^{-8}$~\cite{Belle:2021ysv},
large $\r_{\tau\mu}$ means small $\rtt$.
Even more surprising~\cite{Hou:2021sfl} is that the collider search for 
$g g \to H, A \to \tau \mu$ provides a much better constraint.
Combining both flavor and LHC constraints, 
$\rtt$ seems limited to small values $\rtt\lesssim 0.1$, 
which is small compared with $\l_t \cong 1$.

Small $\rtt$ and $\cg$ will suppress the two-loop contributions to $\ell$EDM:
the top-loop diagrams\ by small $\rtt$,
and $W$-loop diagrams by $\cg\to 0$.
But large $\r_{\tau\mu}(\r_{\mu\tau})$ values can
give rather large $\mu$EDM via the one-loop diagram.

Defining $\r_{\tau\mu}\r_{\mu\tau}=|\r_{\tau\mu}\r_{\mu\tau}|\exp(\phi_{CP})$,
we illustrate in Fig.~\ref{fig: amu-dmu} the range of $d_\mu$ accessible in 
the $|\rho_{\tau\mu}\rho_{\tau\mu}|$-$\phi_{CP}$ plane 
while simultaneously accounting for muon $g-2$, Eq.~\eqref{eq: del_a_mu}.
The left plot is for $m_H=$ 300\;GeV and $m_A = $ 340\;GeV,
which are not far from degenerate, while the right plot is 
for the same $m_H$ but a much heavier $m_A = $ 500\;GeV.
We see from both plots that, given the complexity of $\rho_{\tau\mu}\rho_{\mu\tau}$, 
the solution space for muon $g-2$ projects large $d_\mu$
values that are well within PSI sensitivity!
In the more optimistic scenario where $m_H$ and $m_A$ are farther apart, 
the parameter space in the right plot may even appear to be 
within reach of Fermilab and J-PARC muon g-2 experiments,
namely $\sim 10^{-21}\;e$\,cm.
We caution, however, that this
would demand {\it very} large $|\rho_{\tau\mu}| \sim |\rho_{\mu\tau}|$ values.

On the other hand, one does not expect such extraordinary enhancements for $d_\tau$.
In fact, if muon $g-2$ arises from g2HDM as described,
then  $d_\tau$ will be much smaller than our estimates in Sec.~\ref{sec: III}.
This is because the two-loop contribution to $d_\tau$ in Eq.~\eqref{eq: dtau} 
becomes suppressed by small $\rtt$ and $\cg$, as argued previously.
Furthermore, one-loop enhancement similar to $\mu$EDM is not possible 
because simultaneously large $\r_{\tau\tau}$ and $\r_{\tau\mu}(\r_{\mu\tau})$ 
would make $\tau\to\mu\gamma$ too large at one-loop and violate the Belle bound, 
as we showed recently in Ref.~\cite{Hou:2021qmf}.

What about $e$EDM? 
The leading one-loop results ($\tau$ in the loop) for $e$EDM are
given by Eq.~\eqref{eq: 1loop}, replacing $``\mu"\to ``e"$. 
Therefore, at one loop,
$d_e$ is proportional to $\operatorname{Im}(\r_{\tau e}\r_{e\tau})$.
The $\mu\to e\gamma$ process constrains $\rho_{\tau e}\,(\rho_{\tau e})$ severely.
For values of $\r_{\tau\mu}$ consistent with muon $g-2$ solution,
the current MEG~\cite{MEG:2016leq} bound of
${\cal B}(\mu\to e\gamma) < 4.2 \times 10^{-13}$
gives $|\r_{\tau e}|=|\r_{e\tau}|\lesssim \o(\l_e)$~\cite{Hou:2021qmf},
which is in line with our working assumption in Eq.~\eqref{eq: rho}.
Taking $|\operatorname{Im}(\r_{\tau e}\r_{e\tau})|\simeq \l_e^2$,
$\cg=0$, and $m_H=$ 300~GeV and $m_A =$ 500~GeV,
we find $|d_e|\lesssim 5\times 10^{-32}\;e$\,cm,
which is more than two orders of magnitude below 
the current ACME bound, $|d_e| < 1.1 \times 10^{-29}\;e$\,cm.
For $m_H$ closer to $m_A$, $d_e$ will be further suppressed.
The two-loop results for $d_e$  can be obtained by
replacing $``\mu" \to ``e"$ in the corresponding expressions for
$\mu$EDM in Sec.~\ref{sec: II}. If one assumes $\r_{ee}\sim 0$,
then two-loop diagrams for $e$EDM is essentially given by the
second term of Eq.~\eqref{eq: dmu-mix} (after obvious change of lepton indices),
which depends on $|s_{2\g} \operatorname{Im}\rtt|$ and $m_H$.
For $m_H=$ 300~GeV, we find
$|d_e|\simeq 2 \times 10^{-27} |s_{2\g}\operatorname{Im} \rtt|\;e$\,cm.
With both parameters constrained to be small, $\cg < 0.005$ and $|\rtt|\lesssim0.1$,
it is rather easy to satisfy the current ACME bound,
without resort to the cancellation mechanism~\cite{Fuyuto:2019svr}
to evade the experimental bound. 
It remains to be seen whether parameter space exists 
where simultaneously small $\rtt$ and $\cg$
can still provide solutions to BAU in g2HDM.

\section{Conclusion}\label{sec: V}

We study one- and two-loop mechanisms for charged lepton EDM in g2HDM, 
the two Higgs doublet model that possesses extra Yukawa couplings. 
We find both $d_\mu$ and $d_\tau$ can be enhanced to considerably larger values
 --- thanks to the extra top Yukawa coupling $\rtt\sim \o(1)$.  

The question for g2HDM is how it has managed to hide so well
from scrutiny, if the exotics Higgs bosons are sub-TeV in mass.
This we have elucidated in a previous work~\cite{Hou:2020itz},
where we developed Eq.~\eqref{eq: rho} as 
our working assumption for the strength of extra $\r^f_{ij}$ couplings,
namely $\rho_{ii} \lesssim \o(\lambda_i)$,
 $\rho_{1i} \lesssim \o(\lambda_1)$, and
 $\rho_{3j} \lesssim \o(\lambda_3)$ for $j \neq 1$.
It reflects the mass-mixing hierarchies observed in the SM sector~\cite{Hou:1991un},
and finds further support from the cancellation mechanism~\cite{Fuyuto:2019svr}, 
or interplay between $\rho_{ee}$ and $\rho_{tt}$ 
to satisfy the recent ACME 2018 bound on $d_e$,
while $\rho_{tt}$ still drives baryogenesis.
Flavor changing couplings of $h(125)$ are further suppressed
by the recent emergent phenomenon of alignment, i.e. small $h$--$H$ mixing.

With our working assumption as above and taking central values, 
we showed that $d_\mu$ lies in the ballpark range of $10^{-24}$\;$e$\,cm,
while $d_\tau$ is in the $10^{-23}$\;$e$\,cm range.
These numbers could go up by another order of magnitude
as allowed by Eq.~\eqref{eq: rho}.
But though they are much larger than typical expectations in MFV-like scenarios, 
with future sensitivities as stated in Table~1, it is unlikely 
that $d_\mu$ and $d_\tau$ would be observed in the near future.
Thus, Eq.~\eqref{eq: rho} continues to ``protect'' g2HDM from
revealing itself, and we have advocated elsewhere for 
LHC direct search~\cite{Hou:2020chc}, 
as well as several flavor probes~\cite{Hou:2020itz}.

The recent experimental affirmation of the muon $g-2$ ``anomaly'',
however, brings in some hope, but at a cost.
In g2HDM, the one-loop mechanism can quite simply account for the discrepancy, 
{\it if} $|\rho_{\tau\mu}| \simeq |\rho_{\mu\tau}| = \o(20\lambda_\tau) \sim 0.2$.
This would grossly violate our working assumption of Eq.~\eqref{eq: rho}; 
but {\it Nature} herself is the judge and the jury, and maybe this is what
she has been telling us through the ``anomaly''.
The point is, the same one-loop diagram can generate $d_\mu$
through the complexity of $\rho_{\tau\mu} \rho_{\mu\tau}$,
which {\it can} be tested with projected experimental sensitivities.
But this single-enhanced coupling scenario would then imply 
quite small $\rtt\lesssim 0.1$, while alignment must be near-perfect,
which by themselves are rather peculiar.
These in turn suggest that $d_\tau$ will be rather suppressed,
and $d_e$ can be much below ACME bound 
without resort to cancellation mechanism.

\vskip0.2cm
\noindent{\bf Acknowledgments} 
This research is supported by 
MOST 109-2112-M-002-015-MY3, 110-2639-M-002-002-ASP and 
110-2811-M-002-620 of Taiwan, 
and NTU grants 110L104019 and 110L892101.


\appendix
\begin{widetext}
\section{}{\label{app: fun}}
Below we provide Barr-Zee formulae related to $d_\mu^{\phi Z}$ and charged Higgs loop,
together with expressions for all the loop functions used for calculation of EDMs in this paper.

\subsection{More Barr-Zee formulae}

The Barr-Zee formulae for $d_\mu^{\phi Z}$  are as follows.
Similar to $d_\mu^{\phi \g}$, we decompose the top-loop contributions to $d_\mu^{\phi Z}$ as
$(d_\mu^{\phi Z})_t=(d_\mu^{\phi Z})_{t}^{\rm mix}+ (d_\mu^{\phi Z})_{t}^{\rm extra}$.
The expressions for each term are,
\begin{align}\label{eq: dmuZ-mix}
		\left(d_\mu^{\phi Z}\right)_{t}^{\rm mix}
		=\frac{e\, \alpha\,  s_{2\g} }{256 \pi^3 m_ts_W^2 c_W^2}(1-4\,s^2_W)(1-\tfrac{8}{3}s^2_W)
		\left\{\left[\l_t\operatorname{Im}\r_{\mu\mu}\left(\frac{m_h^2 f(x_{th})-m_Z^2 f(x_{tZ})}{m_h^2-m_Z^2}
															-h\to H\right)\right]\right.\nonumber\\
		+ \left.\biggl[\l_\mu\operatorname{Im}\r_{tt} \,\biggl( f\to g \biggr)\biggr]\right\},
\end{align}
\begin{align}\label{eq: dmuZ-extra}
		\left(d_\mu^{\phi Z}\right)_{t}^{\rm extra}
		=\frac{e\, \alpha  }{128 \pi^3 m_t\,s_W^2 c_W^2}(1-4\,s^2_W)(1-\tfrac{8}{3}s^2_W)
		 \left\{\left[\operatorname{Im}\r_{\mu\mu}\operatorname{Re}\r_{tt}
		 \left(\cg^2 \frac{m_h^2 f(x_{th})-m_Z^2 f(x_{tZ})}{m_h^2-m_Z^2}\right.\right.\right.\nonumber \\
		\left.\left.\left.+ s_\g^2  \frac{m_H^2 f(x_{tH})-m_Z^2 f(x_{tZ})}{m_H^2-m_Z^2}
		+ \frac{m_A^2 g(x_{tA})-m_Z^2 g(x_{tZ})}{m_A^2-m_Z^2}\right)\right]\right.\nonumber \\
		+ \left.\biggl[ \operatorname{Im}\r_{tt}\operatorname{Re}\r_{\mu\mu} \,\biggl( f\leftrightarrow g \biggr) \biggr]\right\},
	\end{align}
The $W$-loop contribution is,
\begin{align}\label{eq: dmuZ-W}
		\left(d_\mu^{\phi Z}\right)_{W}
		=-\frac{e\, \alpha s_\g c_\g (1-4s_W^2)}{128\sqrt{2}\pi^3 v\,s_W^2}
		\operatorname{Im}\r_{\mu\mu} \Bigl[ I_Z(m_h)-I_Z(m_H)\Bigr],
	\end{align}
and lastly, the charged Higgs loop diagram gives,
\begin{align}\label{eq: dmuZ-H+}
		\left(d_\mu^{\phi Z}\right)_{H^+}
		=-\frac{e\, \alpha\, (c_W^2 -s_W^2)(1-4s_W^2)}{256\sqrt{2}\pi^3 v\,s_W^2 c_W^2}
		\operatorname{Im}\r_{\mu\mu}\Bigl[c_\g C_{hH^+H^-} I_3^Z(m_H^+,m_h)
		-s_\g C_{HH^+H^-} I_3^Z(m_H^+,m_H)\Bigr],
	\end{align}
where the trilinear couplings $\phi H^+H^-$ are defined as
$-i v C_{\phi H^+ H^-}$ in notation of Refs.~\cite{Gunion:2002zf,Davidson:2010xv}.
Note that $C_{A H^+H^-} = 0$.

The charged Higgs loop contribution for $d_\mu^{\phi \g}$ and $d_\mu^{H^\pm W^\pm}$ are,
\begin{align}\label{eq: dmu-H+loop}
		\left(d_\mu^{\phi \g}\right)_{H^+}
		=-\frac{e\, \alpha}{32\sqrt{2}\pi^3 v}
		\operatorname{Im}\r_{\mu\mu}
		 \Bigl[c_\g C_{hH^+H^-} I_3^\g(m_H^+,m_h)
		 -s_\g C_{HH^+H^-} I_3^\g(m_H^+,m_H)\Bigr],
\end{align}
\begin{align}\label{eq: dmu-HW-H+loop}
		\left(d_\mu^{H^+ W^+}\right)_{H^+}
		=-\frac{e}{256\sqrt{2}\pi^4 v}
		\operatorname{Im}\r_{\mu\mu}
		\Bigl[c_\g C_{hH^+H^-} {\cal I}_5(m_h, m_H^+,)
		-s_\g C_{HH^+H^-} {\cal I}_5(m_H, m_H^+,)\Bigr].
\end{align}

\subsection{Loop Functions}
The loop functions appearing in various Barr-Zee formulae of lepton EDM are listed below.

The functions appearing in top quark contribution to $d_\mu^{\phi \g}$ and $d_\mu^{\phi Z}$ are~\cite{Barr:1990vd},
	\begin{align}
		f(a) = a\int_0^1 dz \frac{1/2-z+z^2}{z(1-z)-a}\log\frac{z(1-z)}{a},\quad 
		g(a) = \frac{a}{2}\int_0^1 dz \frac{1}{z(1-z)-a}\log\frac{z(1-z)}{a}.
	\end{align}

The functions appearing in $W$-loop contribution to $d_\mu^{\phi \g}$ and $d_\mu^{\phi Z}$ are~\cite{Abe:2013qla},	
\begin{align}
		I_{V}(m_\phi) = \frac{m_W^2}{m_\phi^2 -m_V^2}\left\{\left[6 - \frac{m_V^2}{m_W^2}
		+\left(1-\frac{m_V^2}{2m_W^2}\right)\frac{m_\phi^2}{m_W^2}\right]\right.
		\left(\frac{m_\phi^2}{m_W^2}f(m_W^2/m_\phi^2)-\frac{m_V^2}{m_W^2}f(m_W^2/m_V^2)\right)\nonumber\\
		 -\left[-10 + \frac{3m_V^2}{m_W^2}+\left(1-\frac{m_V^2}{2m_W^2}\right)\frac{m_\phi^2}{m_W^2}\right]
		\left.\left(\frac{m_\phi^2}{m_W^2}g(m_W^2/m_\phi^2)-\frac{m_V^2}{m_W^2}g(m_W^2/m_V^2)\right)\right\}.
\end{align}

The functions appearing in $H^+$-loop contribution to $d_\mu^{\phi \g}$ and $d_\mu^{\phi Z}$ are~\cite{Abe:2013qla},
\begin{align}
		I_{3}^V(m_H^+, m_\phi) = \frac{v^2}{m_\phi^2 -m_V^2}
		 \left\{\left[\frac{m_V^2}{m_{H^+}^2}\Bigl(f(m_{H^+}^2/m_V^2)
		 -g(m_{H^+}^2/m_V^2)\Bigr)\right]
		-\biggl[V \to \phi\biggr]\right\}.
\end{align}

The functions appearing in fermion loop contributions to $d_\mu^{H^+ W^+}$ are~\cite{Bowser-Chao:1997kjp,Jung:2013hka},
\begin{align}
		F_q(a, b) = \frac{T_q(a)-T_q(b)}{a-b};~~q=t, b,
\end{align}
where
\begin{align}
		T_b(a) &= \frac{1-3a}{a^2}\frac{\pi^2}{6}+\left(\frac{1}{a}-\frac{5}{2}\right)\log a
		- \frac{1}{a}-\left(2-\frac{1}{a}\right)\left(1-\frac{1}{a}\right){\rm Li}_2(1-a),\\
		T_b(a) &= \frac{2a-1}{a^2}\frac{\pi^2}{6}+\left(\frac{3}{2}-\frac{1}{a}\right)\log a
		+ \frac{1}{a}-\frac{1}{a}\left(2-\frac{1}{a}\right){\rm Li}_2(1-a).
\end{align}

Finally, the functions appearing in diagrams with $W^+$ and $H^+$ loops
for  $d_\mu^{H^+ W^+}$ are~\cite{Abe:2013qla},
\begin{align}
	{\cal I}_{4(5)}(m_\phi, m_{H^+})
	= \frac{m_W^2}{m_{H^+}^2-m_W^2}	\Bigl[I_{4(5)}(m_W, m_\phi)-I_{4(5)}(m_{H^+},m_\phi,)\Bigr],
\end{align}
where,
\begin{align}
	I_4(m_i,m_\phi)&=
	\int_0^1dz
	\frac{m_i^2\left(z(1-z)^2-4(1-z)^2+\frac{m_{H^+}^2-m_\phi^2}{m_W^2}z(1-z)^2\right)}{m_W^2(1-z)+m_\phi^2 z-m_i^2 z(1-z)}\log\left(\frac{m_W^2(1-z)+m_\phi^2 z}{m_i^2 z(1-z)}\right),\\
	I_5(m_i,m_\phi)&=
	2\int_0^1dz\frac{m_i^2 z(1-z)^2}{m_{H^+}^2(1-z)+m_\phi^2 z-m_i^2 z(1-z)}
	\log\left(\frac{m_{H^+}^2(1-z)+m_\phi^2 z}{m_i^2 z(1-z)}\right).
\end{align}

\end{widetext}

\end{document}